\documentclass[12pt]{article}
\usepackage[utf8]{inputenc}
\usepackage[margin=1.0in]{geometry}

\usepackage{float}
\usepackage{amsmath}
\numberwithin{equation}{section}
\usepackage{amsmath,latexsym}
\usepackage{amssymb}
\usepackage{graphicx}
\usepackage{mathtools}
\usepackage{multicol}
\usepackage{subcaption}
\usepackage[T1]{fontenc}
\usepackage[latin9]{luainputenc}
\usepackage{babel}

\pdfoutput=1
\usepackage{hyperref}
\usepackage{pdfpages}
\usepackage{microtype}

\title{Back and Forth with Akito Arima}
\author{Larry Zamick and Castaly Fan\\
Rutgers, the State University of New Jersey, Piscataway, NJ 08854
}

\begin{document}

\maketitle
\noindent \textbf{DOI:} \url{https://doi.org/10.1142/9789819813353_0020}

    \begin{abstract}
    In 1967 Akito Arima spent a year as a visiting professor in the physics department of Rutgers University. In this work we pay tribute to him by discussing topics that we worked on that were directly influenced buy his works or were closely related to his interests. These include nuclear Symmetries, magnetic and other moments, analytic expressions in the single $j$ shell model and schematic interactions.
    \end{abstract}

\tableofcontents

\clearpage

\section{Introduction}

    Akito Arima was a visiting professor at Rutgers in 1967. He was not the only Japanese visitor. Also Shiro Yoshida and a bit later a very bright student of Arima-Koichi Yazaki. Yoshida stayed on as a professor for about 5 years and I wrote the following article with him. ``Electromagnetic Moments and Transitions Annual Review of Nuclear Science - Vol 22''. I developed lifelong friendships with all these people. I was invite to give one of the banquet speeches at a 1972 conference in Japan honoring Akito. I said diplomatically: ``Akito Arima is the most respected nuclear theorist in the world and Shiro Yoshida is the most respected nuclear theorist in Japan''.

    Akito's work was not only a big influence on the topics that I chose but was also of importance to the experimental group at Rutgers - Noemie Koller, Gerfried Kumbartzki, and from Bonn, Karl Heinz Speidel. This group measured magnetic moments of excited states of even-even nuclei and the work of Arima and Hori, well known at that time, provided a solid theoretical background.

    When I was postdoc at Princeton Gerry Brown suggested to me and to a then student Harry Mavromatis that we also work on magnetic moments. The Arima-Horie theory was basically first order perturbation theory so we were to do second order. This is important for a closed major shell plus (or minus) one nucleon because for such cases first order vanishes. Ichimura and Yazaki also had worked on this. By the way soon after Gerry Brown left Princeton for Stony Brook an the took Akito with him. I recall at conferences Akito would stand up to ask a question: He started with Akito Arima, Stony Brook and Tokyo and then the fearsome question.

    Besides magnetic moments I will here discuss topics which paralleled the interest of Arima - at least I hope they did. These include Nuclear Symmetries, quadrupole moments, single $j$ shell properties and schematic interactions. In the same time period - after 2000, Akito Arima and Yu-Min Zhao wrote many papers on the single $j$ shell (e.g.number of states of identical particles ), and I had written a few myself. I recall sending emails to Akito about this but got no replies. I related this to Igal Talmi who laughed and said ``Akito does not answer anyone's emails.'' At the same Akito and Yu-Min were very generous in mentioning my works on these topics and I reciprocated in turn. I hope the next few sections will help to covey how important Arima's presence both at Rutgers and on the world scene helped to enlivened my life in physics.

\section{Symmetries}

    Akito's Arima contributions to the subject of symmetries in nuclei is overwhelming. Here we show some work on this topic which we hope would have met with his approval.

    \subsection{Isospin}
    
    Note in Table \ref{tab:Ca,Sc,Ti_spcetra} that the excitation energies of the even $J$ states in $^{42}$Ca, $^{42}$Sc and $^{42}$Ti are early the same. This is evidence of the charge independence of the nuclear force. The fact that odd $J$ states appear only in $^{42}$Sc shows the Pauli principle in action. In $^{42}$Ca and $^{42}$Ti we have 2 identical nucleons so we can only have antisymmetric states. This tells us that in the $j^2$ configuration states with even $J$ are antisymmetric. In $^{42}$Sc we do not have identical nucleons so we can have symmetric states. These are the odd $J$ states. And the there is the multiplicity rule. Even $J$ states occur 3 times so $(2T+1) = 3$ and so $T=1$. The odd $J$ states occur only once so $(2T+1) =1$ and so $T=0$.
    
    \begin{figure}[H]
        \centering
        \captionsetup{width=0.8\linewidth}
        \includegraphics[width=0.75\textwidth]{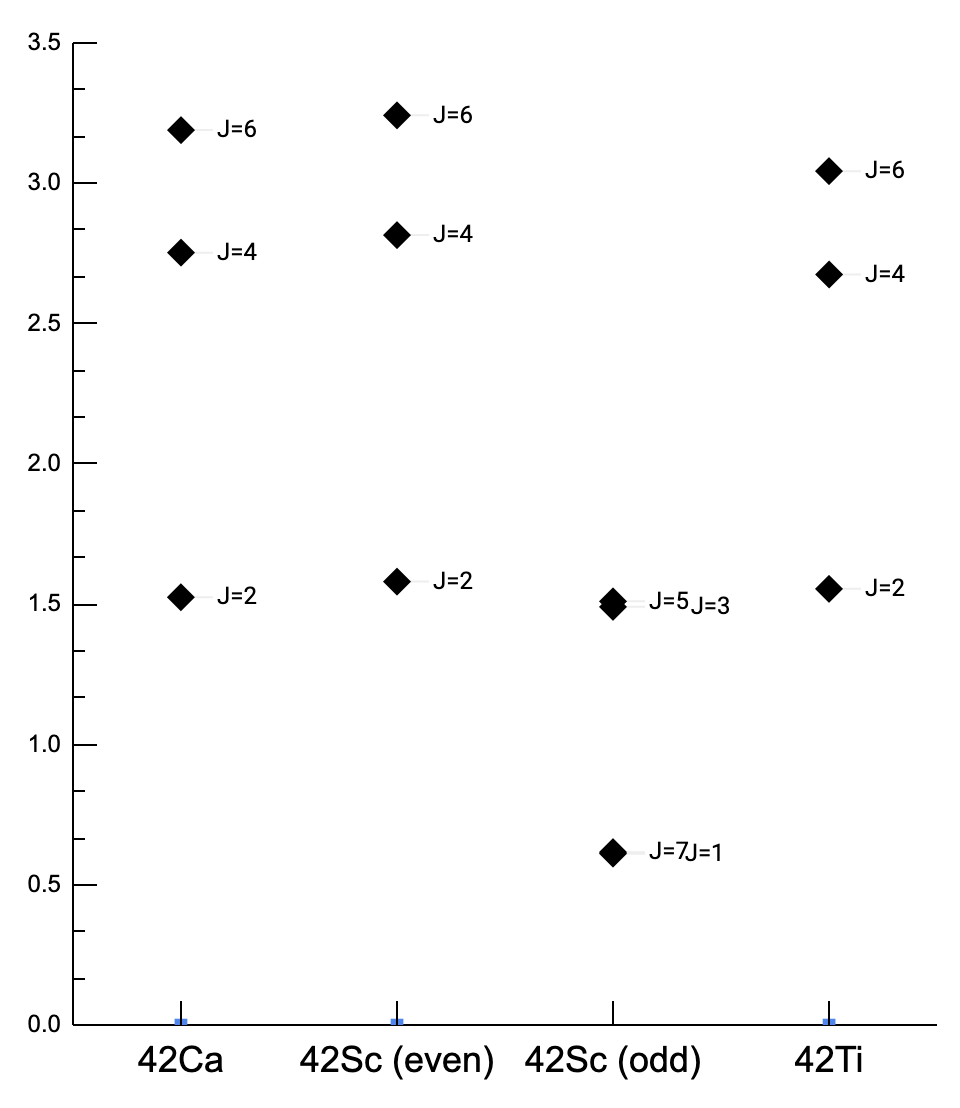}
        \caption{Energy levels of $^{42}$Ca, $^{42}$Sc and $^{42}$Ti shown in order to display the near charge independence of the nuclear force.}
        \label{fig:el}
    \end{figure}
    
     \begin{table}[H]
        \centering
        \caption{The spectra of $^{42}$Ca, $^{42}$Sc, and $^{42}$Ti.}
        {\renewcommand{\arraystretch}{1.5}
        \begin{tabular}{|p{1.5cm}||p{2.5cm}|p{2.5cm}|p{2.5cm}|}
        \hline
        & $^{42}$Ca   & $^{42}$Sc   & $^{42}$Ti   \\ \hline\hline
        $J=0$ & 0.0000 & 0.0000 & 0.0000 \\ \hline
        $J=1$ &  - & 0.6110 &  - \\ \hline
        $J=2$ & 1.5247 & 1.5803 & 1.5546 \\ \hline
        $J=3$ & - & 1.4904 & -  \\ \hline
        $J=4$ & 2.7524 & 2.8153 & 2.6746 \\ \hline
        $J=5$ & -  & 1.5100 & -  \\ \hline
        $J=6$ & 3.1893 & 3.2420 & 3.0430 \\ \hline
        $J=7$ & -  & 0.6163 &  - \\ \hline
        \end{tabular}}
    \label{tab:Ca,Sc,Ti_spcetra}
    \end{table}

    \subsection{Signature}
    
    In the f$_{7/2}$ paper of McCullen et al. \cite{01}\cite{02} we also discuss briefly signature selection rules. For say $^{48}$Ti we have a system of 2 protons and 2 neutron holes in the f$_{7/2}$ shell. We find that the wave functions are either even or odd under the interchange of protons and neutron holes. This is different from isospin in that a state of even signature and one of odd signature can have the same isospin.

    It has been shown that this signature property leads to several selection rules. For example, for the electric quadrupole operator the B(E2) between states of opposite signature is proportional to (ep + en),and between states of the same signature to (ep - en). The quadrupole moment of the $2^{+}$ state is proportional to (ep-en). It turns out that the $2+i$ state of Ti in the single $j$ shell calculation has odd signature, but the 2+2 state has even signature. Hence B(E2) from the $J=0$ ground state to 2(1) goes as (ep + en)$^2$ whilst to 2(2) as (ep - en)$^2$.

    Consider next the double Gamow-Teller operator $(\sigma t_{-})_{i} (\sigma t_{-})_{j}$ connecting $^{48}$Ca to $^{48}$Ti. Zamick and Moya de Guerra \cite{03} showed that in this single $j$ shell model the transition to the 2(1) state (negative signature) vanishes because of the signature selection rule. On the other hand a transition to the 2(2) (positive signature) state is allowed. This simple model shows that there might be surprises when calculating double beta decay transitions.

    \subsection{Seniority}
    
    Some of the well-known statements and theorems concerning states of good seniority are

        \begin{enumerate}
            \item The seniority is roughly the number of identical particles not coupled to zero. Hence, for a single nucleon the seniority $v$ is equal to 1. For two nucleons in a $J = 0$ state we have $v = 0$, but for $J = 2, 4, 6$, etc., $v = 2$. For three nucleons there is one state with seniority $v = 1$, which must have $J = j$; all other states have seniority $v = 3$.

            \item The number of seniority-violating interactions is $[(2j - 3)/6]$, where the square brackets mean the largest integer contained therein. For $j = 7/2$ there are no seniority-violating interactions, while for $j = 9/2$ there is one.

            \item With seniority-conserving interactions, the spectra of states of the same seniority is independent of the particle number.

            \item At midshell we cannot have any mixing of states with seniorities $v$ and $v + 2$; one can mix $v$ and $v + 4$ states.
        \end{enumerate}

    \begin{figure}[H]
        \centering
        \captionsetup{width=0.9\linewidth}
        \caption{We quote from the paper \cite{seniority}:}
        \includegraphics[width=0.75\textwidth]{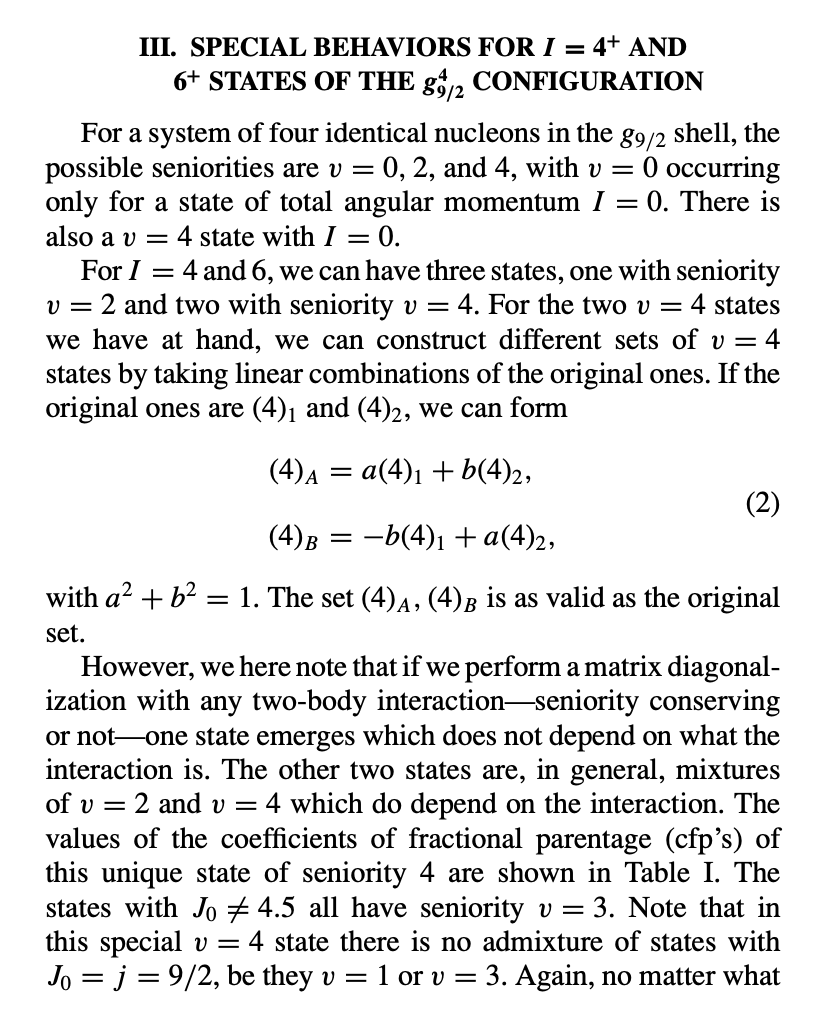}
        \label{fig:s1}
    \end{figure}
    
    \begin{figure}[H]
        \centering
        \includegraphics[width=0.75\textwidth]{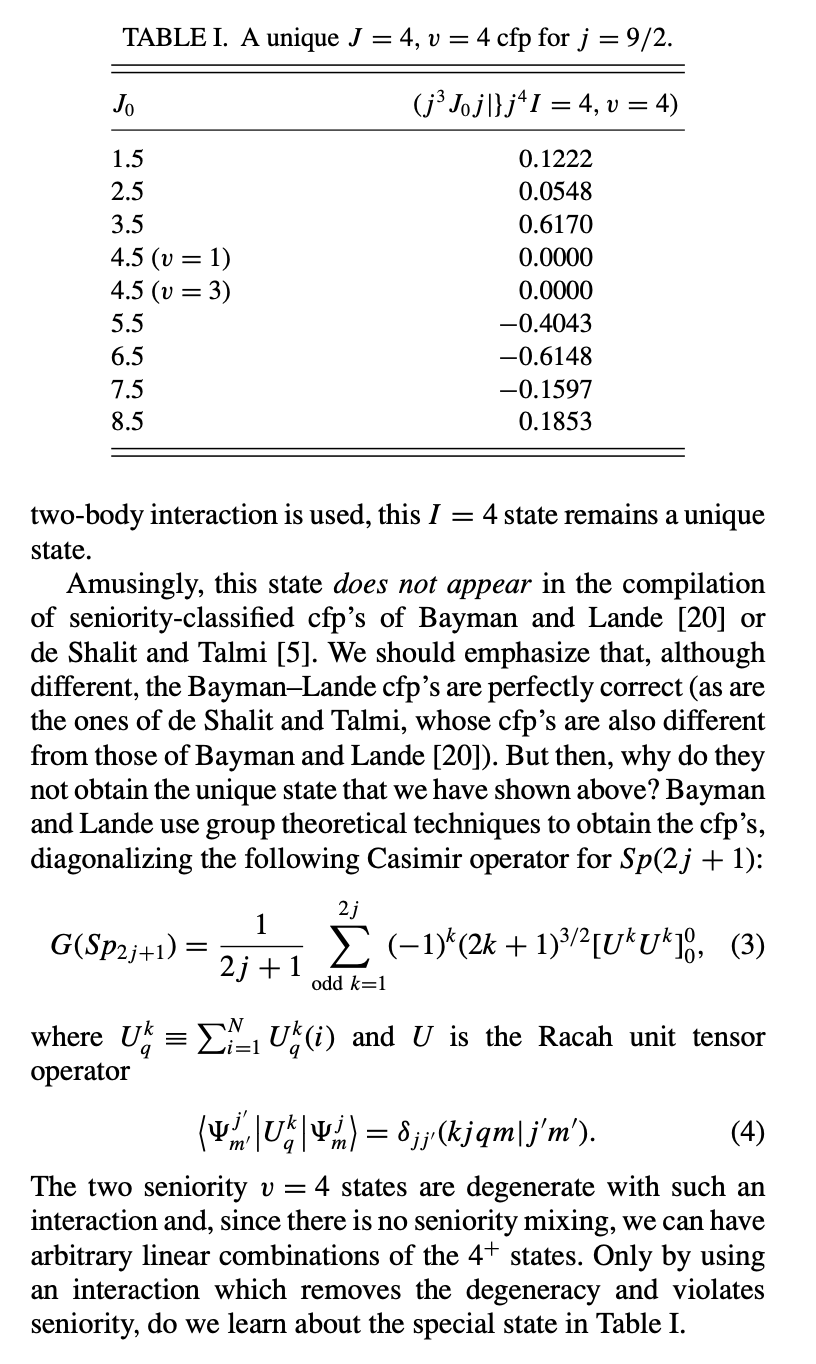}
        \label{fig:s2}
    \end{figure}

\section{Nuclear Moments}

    \subsection{Magnetic moments}
    
    With $n$ nucleons of one kind there are simple formulas for nuclear moments in a single $j$ shell. For example all $g$ factors should be the same. From this it follows that for states of the same $J$ the magnetic moments should be the same. The magnetic moment of a free neutron (in units of nuclear magnetons) is $\mu_{n} = -1.913$ and that of a free proton is $\mu_{p} =+2.793$. In a single $j$ shell of neutrons with $j=l +1/2$ the magnetic moments are predicted to be the same as those of a free neutron-namely $-1.913$ ; for protons it is $(2.793 +l) (\mu_{n})$. Here $L$ is the orbital angular momentum.
    
    The single particle magnetic moments, commonly called the Schmidt moments are given here: 
    \begin{enumerate}
        \item for an odd proton: 
        \begin{itemize}
            \item $\mu = j-1/2 + \mu_{p}$ for $j = l + 1/2$ 
            \item $\mu = j /(j + 1) [ j + 3/2 - \mu_{p} ]$ for $j = l - 1/2$ 
        \end{itemize}
        \item for an odd neutron:
        \begin{itemize}
            \item $\mu = \mu_{n}$ for $j = l + 1/2$
            \item $\mu = - j/ ( j+1) \mu_{n}$ for $j =l-1/2$. 
        \end{itemize}
    \end{enumerate}    
    We can discuss these in a more physical manner. The magnetic moment of a free neutron (in units of nuclear magnetons) is $\mu_{n} = -1.913$ and that of a free proton is $\mu_{p} =+2.793$. In a single $j$ shell of $n$ neutrons with $j=l +1/2$ the magnetic moments are predicted to be the same as those of a free neutron-namely $-1.913$; for $n$ protons it is $(2.793 +l)$. Here $l$ is the orbital angular momentum. For a $j=l-1/2$ neutron we have a quantum effect so that the magnetic moment is only minus that of a free neutron in the large $j$ limit. In general it is $-j / (j+1)$ that of a free neutron.

\begin{figure}[H]
        \centering
        \captionsetup{width=0.8\linewidth}
        \includegraphics[width=0.65\textwidth]{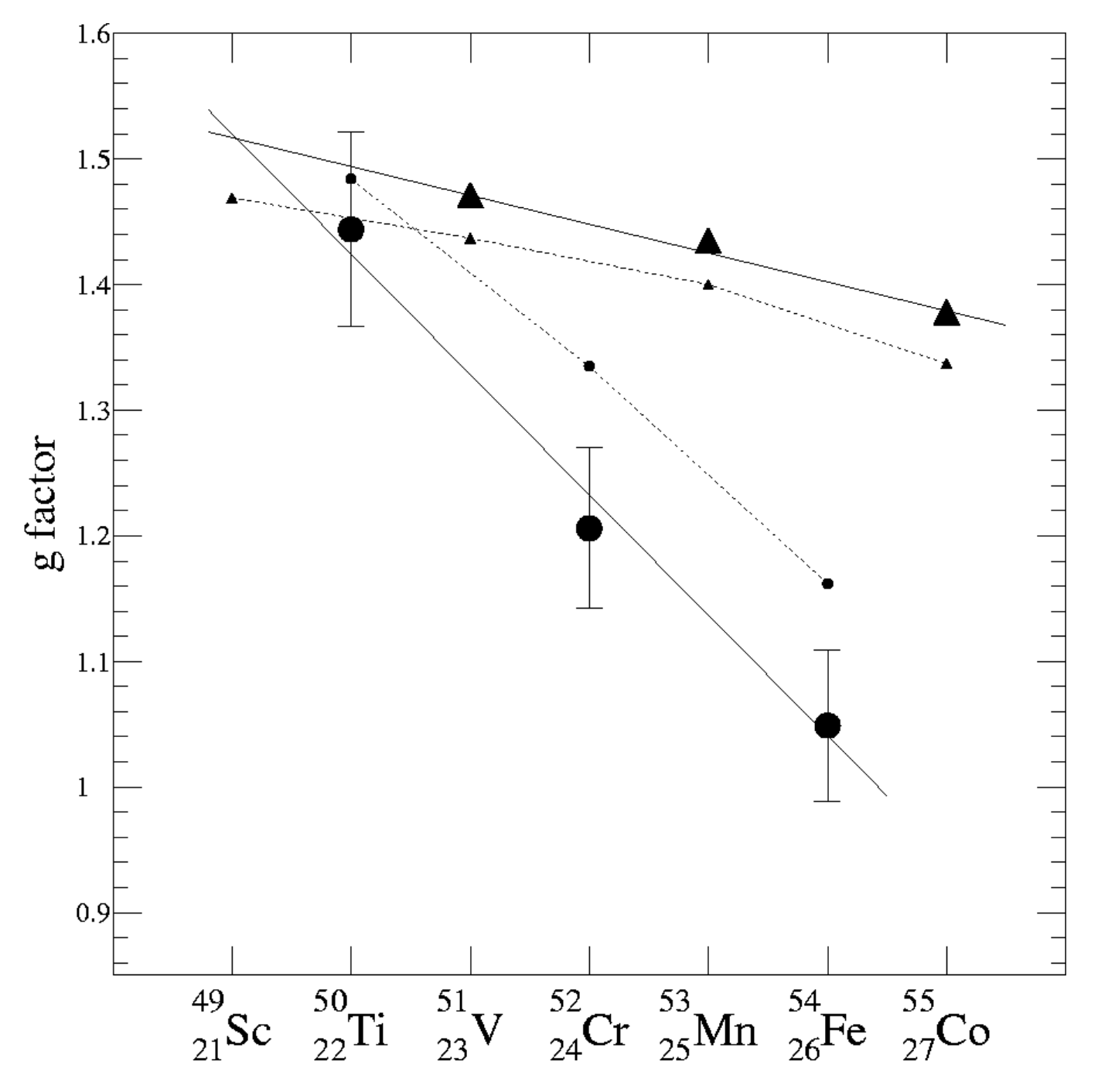}
        \caption{Triangles: factors of ground states of odd isotopes. Circles: g factors of $2^{+}$ states of even isotopes. Light lines: theory.}
        \label{fig:oei}
    \end{figure}

    In the sixties Arima was already famous for the Arima-Horie theory for quenching magnetic moments which is basically first order perturbation theory \cite{x}\cite{y}. I am showing a figure that I like (Figure \ref{fig:oei}) because it has both theorists and experimentalists at Rutgers and Bonn testing out the Arima-Horie theory of quenching.

    \subsubsection{Magnetic moments of isotones and isotopes}

\begin{figure}[H]
        \centering
        \captionsetup{width=0.8\linewidth}
        \includegraphics[width=0.65\textwidth]{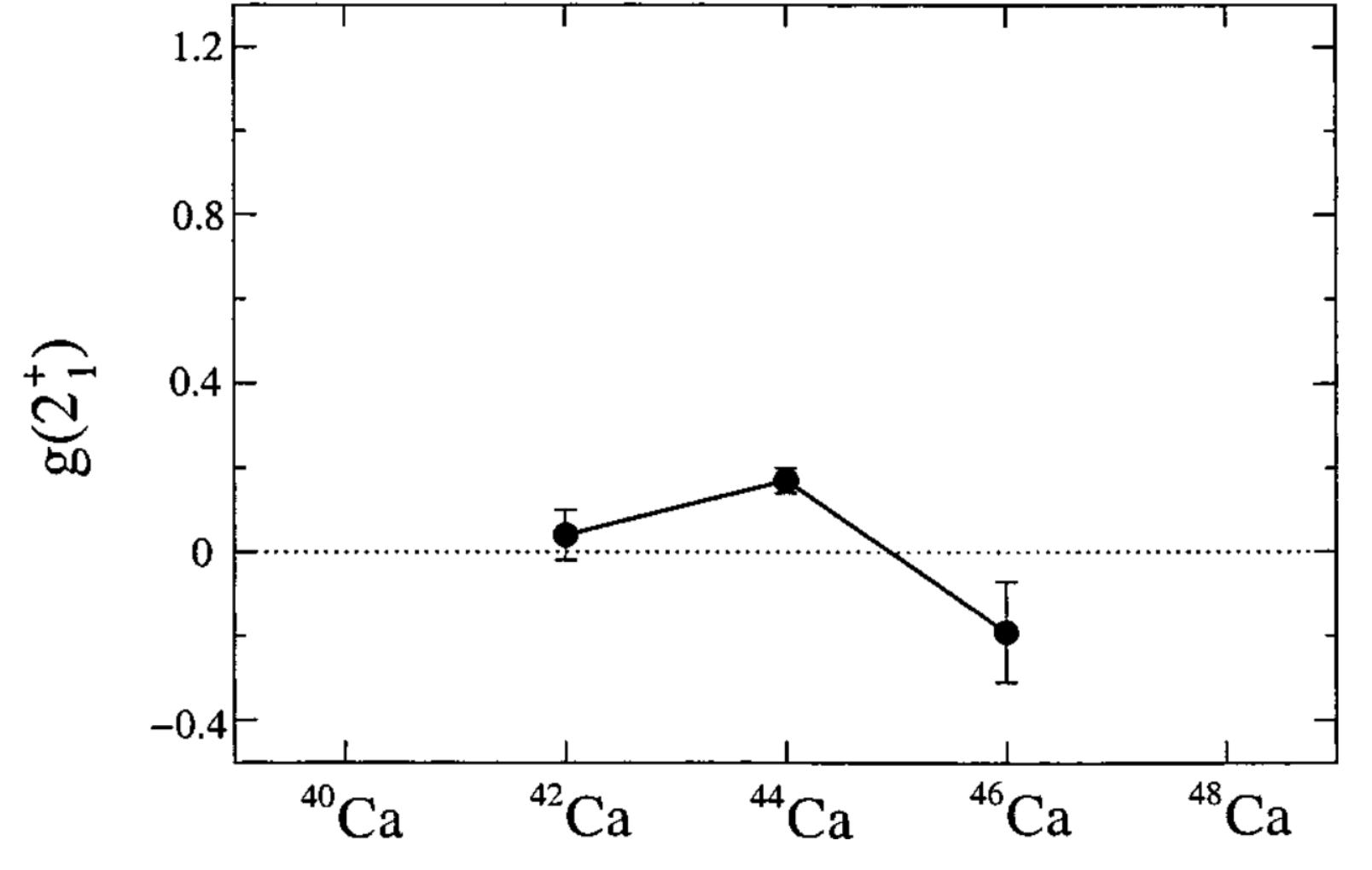}
        \caption{Source: Phys. Rev C 68, 061302(R) (2003).}
        \label{fig:Cg}
    \end{figure}

    In the single shell model, all factors (magnetic moment /angular momentum) are the same. Many people are under the impression that in Arima-Horie that is also true of the quenched g factors. But that is not the case. Rather they are predicted to lie on a straight line with a negative slope. Experimental confirmation of this is shown beautifully in the figure. The slope for states of even nuclei is different than for the ground states of odd nuclei. This is at it should be. One might think one could also apply this to the Calcium isotopes but there intrude states come in to spoil the picture, especially of the g factors of the $2^{+}$ states \cite{z1}\cite{z2}. For example, for the state of $^{44}$Ca the g factor in the f$_{7/2}$ model is about $-0.5$ , whilst that of a highly deformed intruder state is about $+0.5$. We explain the measured result of close to zero for this state by assuming a 50\% admixture of the shell model and intruder state (see Figure \ref{fig:Cg}) \cite{z}.

    \subsubsection{Second order perturbation theory}
    
    My first foray into this subject of magnetic moments was actually not on the work above -- i.e. first order perturbation theory. Rather with Gerry Brown and a Princeton student Harry Mavromatis we dealt with cases where first order perturbation theory was zero and we had to go to second order -- much more complicated \cite{z3}\cite{z4}\cite{z5}. Work on second order was also done in Japan by Ichimura and Yazaki \cite{z6}. We deal with a closed major shell plus a nucleon e.g. $^{17}$O, $^{17}$F, $^{41}$Ca, $^{41}$Sc. These calculations involved a lot of complicated Feynman diagrams. The results were in the right direction to remove the discrepancy from theory and experiment. 
    
    The calculations were done first with the Kallio-Koltveit(KK) interaction \cite{KK} which does not contain a central interaction and then with the more realistic Hamada Johnson  (HJ) interaction \cite{HJ}. Note that with KK the corrections for mirror pairs are equal and opposite i.e. there is no isoscalar correction.  However with HJ ,which contains a tensor interaction we do get an isoscalar correction in second order perturbation theory.  This result was proved by the authors. The above interactions   were soon  superseded by the Kuo-Brown matrix elements \cite{kuo}.
    
     \begin{table}[H]
        \centering
        \captionsetup{width=0.8\linewidth}
        \caption{The calculated sums for the two mirror pairs and the results for the second order corrections to the magnetic moments of the eight nuclei \cite{10}.}
        \includegraphics[width=0.95\textwidth]{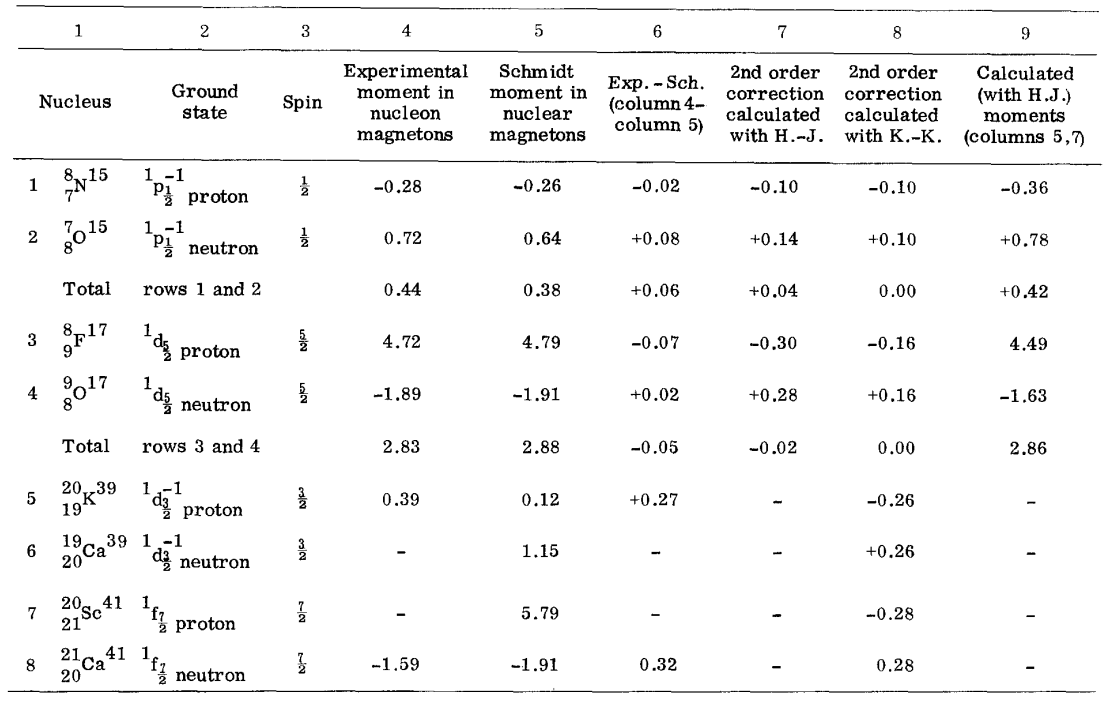}
    \label{tab:nm}
    \end{table}

    \subsubsection{Isoscalar magnetic moments}
    
    Isoscalar magnetic moments have been extensively discussed by S. S. Yeager, L.Zamick, Y.Y. Sharon and S.J.Q. Robinson \cite{04} Isoscalar magnetic moments are much closer to the Schmidt values than the isovector ones. Nevertheless, there are small but systematic deviations. It was noted by Talmi \cite{05} that ``The experimental values of $\langle S \rangle$ seems to follow a simple rule. They are always smaller in absolute value than the values calculated in $jj$ coupling.'' Arima, however, noted \cite{06} that the smallness of the isoscalar deviation is due to the small isoscalar spin coupling (0.44) relative to that of the isovector coupling (2.353). If one divides the deviation by the lowest order result one can get a rather large ratio even in the isoscalar case, even up to 50\%. 
    
     \begin{table}[H]
        \centering
        \captionsetup{width=0.8\linewidth}
        \caption{In Table 2 of the work of Yeager et al \cite{04} we show a table of empirical magnetic moments  as well as shell model calculations and Schmidt estimates. The close agreement of the latter with experiment  should be noted.}
        \includegraphics[width=0.75\textwidth]{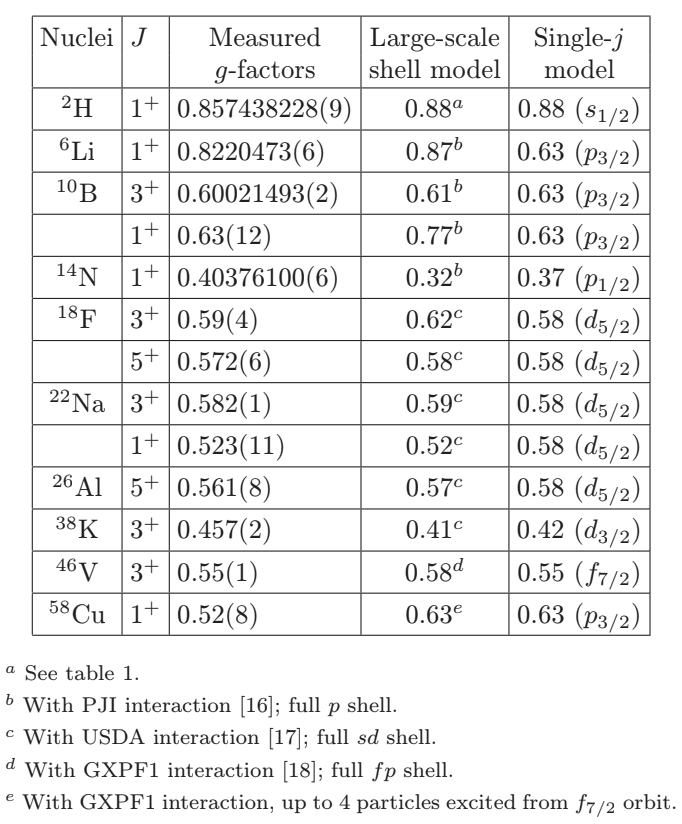}
    \label{tab:gf}
    \end{table}

    \subsection{Quadrupole moments}
    
    For quadrupole moments there is also an $n$ dependent simple formula for ground states of odd nuclei in a single $j$ shell
        \begin{equation*}
            Q(n) =[(2j+1-2n)/(2j-1)] Q(sp)
        \end{equation*}
    Note that for a single hole $n=2 j$. The formula becomes $Q= - Q(sp)$. I.e. the quadruple moment of a hole is minus that of a particle. We can understand this another way. A nuclear moment is the expectation value of a moment operator in a state with $M=J$, 
        \begin{equation*}
             Q^{2}= \langle \Psi^{J}_{J}| Q^{2}_{0} | \Psi^{J}_{J} \rangle.
        \end{equation*}
    To create a hole nucleus in a state with $M=J$ we have to remove a nucleon from a closed shell with $M=-J$. The value of $Q$ for a closed shell is zero-this is the the sum of $Q$ for the hole nucleus and the nucleon removed. The value of $Q^2$ in a state with $M=J$ is the same as it is for $M= -J$ -- namely $Q(sp)$. So we have $Q(\text{hole})+Q(sp)=0$ or $Q(\text{hole})=-Q(sp)$. For magnets moments we have the opposite the value for $-J$ is minus that for $+J$. Thus we have 2 minus signs and $\mu(\text{hole})=\mu(sp)$.

    As a first example of the sturdiness of the shell model we look at the work of Ruiz et al. \cite{1} on measurements and theoretical analysis quadrupole moments of odd A nuclei in the ``f-p'' region. They measured the quadruple moments of the $J=7/2^{-}$ ground states of Calcium isotopes with $A=43,45,47$ which have ground state spins $J=7/2^{-}$. They did not do $A=41$ but this case could be obtained from another source. They also obtained results for $A=49,51$ with $J=3/2^{-}$ spins.

    A starting point for $A=41$ to 47 would be the f$_{7/2}$ shell while for $A=49,51$ it would be the p$_{3/2}$ shell. The theoretical calculations were performs with many interactions and different model spaces. The latter include complete pf space, (pf $+$ g$_{9/2}$) and breaking the $^{40}$Ca core by allowing 2p-2h admixtures. They use effective charges of 1.5 for the protons and 0.5 for the neutrons. In general the calculations are in excellent agreement with the measurements. We will not go into further details about the calculations except to say that they involve an enormous number of shell model configurations.

\begin{figure}[H]
        \centering
        \captionsetup{width=0.8\linewidth}
        \includegraphics[width=0.75\textwidth]{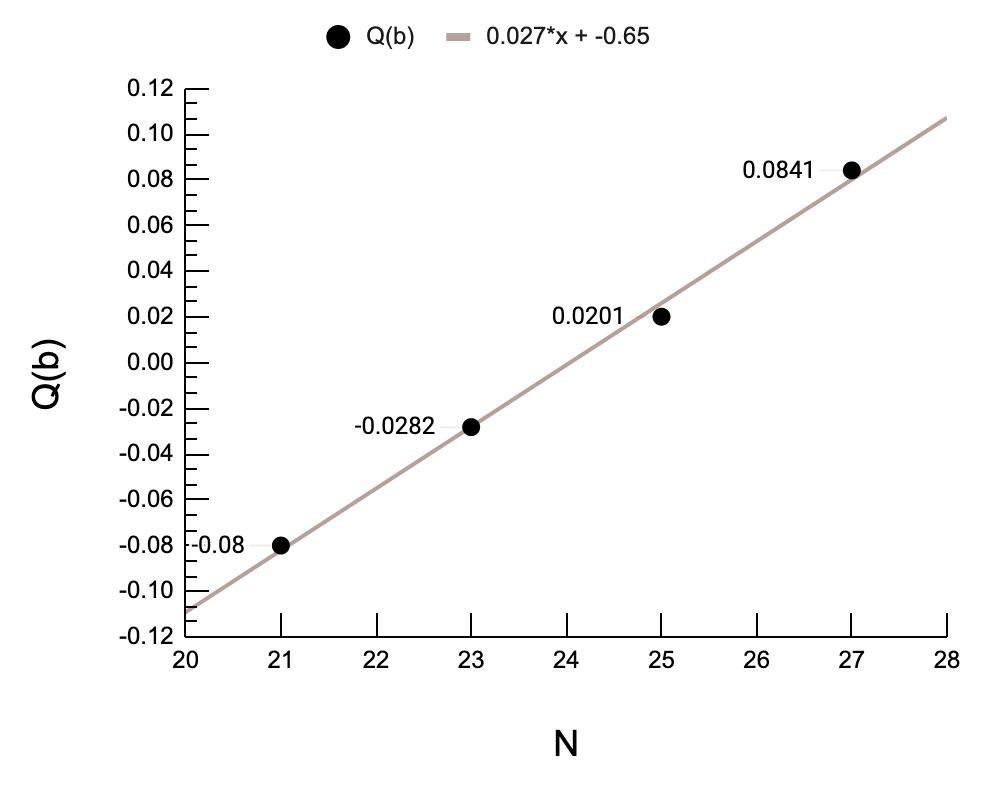}
        \caption{The quadrupole moments vs. $A$ from $A = 41$ to 47.}
        \label{fig:qm}
    \end{figure}

    Rather in Fig \ref{fig:qm} we show the quadrupole moments vs. $A$ and show the the remarkable result that the measured moments from $A=41$ to $A=47$ lie, to an excellent approximation on a straight line. As noted in the introduction this is exactly what a single $j$ calculation predicts. To repeat $Q= (2j-1-2n)/(2j-1)*Q(s.p.)$ This simple result seems to survive the large shell attack. For $A=51$ the measured quadrupole moment $Q=+0.04$ b. It is nearly equal and opposite of that for $A=49$ $Q=-0.04$ b. This is the prediction of the simplest shell model in which $A=49$ consist of a single p$_{3/2}$ neuron and $A= 51$ of a p$_{3/2}$ hole.

    Before leaving this section we should mention that a purist might say that the real prediction of single $j$ is that all the charge quadrupole moments are zero because the neutrons have no charge. We have to assign an effective charge to the neutrons, popular choice being $e_{\text{eff}}=0.5$. But note that even the large space calculations including those of \cite{1} require effective charges in order to get agreement with experiment. In first order perturbation theory the effective charge comes from $\Delta N =2$ excitations. For example for $^{41}$Ca excitations from 0p to 1p; from 0d to 0g, 1d, and 2s. As large as model spaces are in \cite{1} and in nearly all other calculations these configurations are not present and one needs to insert effective charges.
    
    We next consider an ``empirical'' limit for expectation of the isoscalar spin operator $\langle\sigma\rangle$. In the single particle model (i.e. Schmidt) the value for $j=L-1/2$ is $-j /(j+1)$. While the value for$ j=L+1/2$ is one. For the most part the measured values lie between these 2 limits and this has been called an empirical rule. Occasionally some one comes up with an exception. In the work of Kramer et al. \cite{xx}, the magnetic moment of $^{21}$Mg is measured, which when combined with the moment of $^{21}$F yields an isoscalar magnetic moment and an expectation value of the spin operator.

    They find a value of $\langle\sigma\rangle = 1.15(2)$. They call this an anomalous result. We pointed out however that the ``empirical rule'' is not a theoretical rule. In LS coupling the value of the spin operator is
        \begin{equation}
            \langle\sigma\rangle = [S(S + 1) + J (J + 1) - L(L + 1)]/(J + 1).
        \end{equation}
    The smallest value is $-2S J/(J+1)$. And the largest value is 2S. So we can get in principle get values of $\langle\sigma\rangle$ that are greater than one. The empirical rule is not a theoretical rule \cite{11}.

    \subsubsection{Empirical rule}
    
\begin{figure}[H]
        \centering
        \captionsetup{width=0.8\linewidth}
        \includegraphics[width=0.75\textwidth]{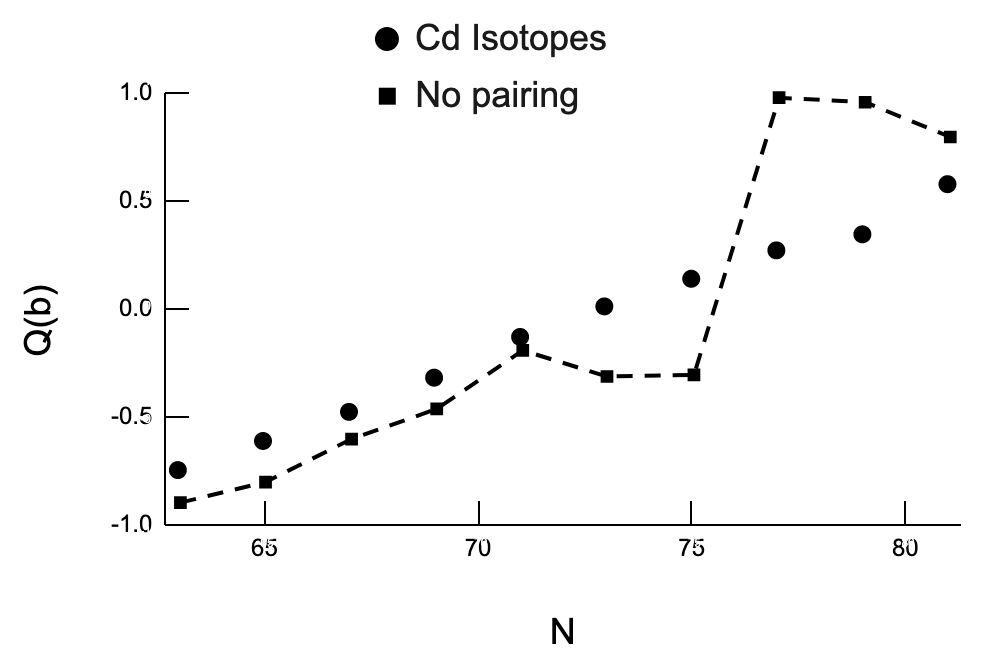}
        \caption{$Q$ vs. $N$.}
        \label{fig:QN}
    \end{figure}

    As shown in Fig \ref{fig:QN} from the work of  P.W. Zhao et al. \cite{pw} we have the strange case where without pairing we get a complicated behavior of $Q$ vs $N$ but when pairing is included one gets a linear behavior as in the single $j$ shell. However the linear curve has more entries than are present for an h$_{11/2}$ shell.  
    
    \subsection{Isotope shifts}
    
\begin{figure}[H]
        \centering
        \captionsetup{width=0.8\linewidth}
        \includegraphics[width=0.75\textwidth]{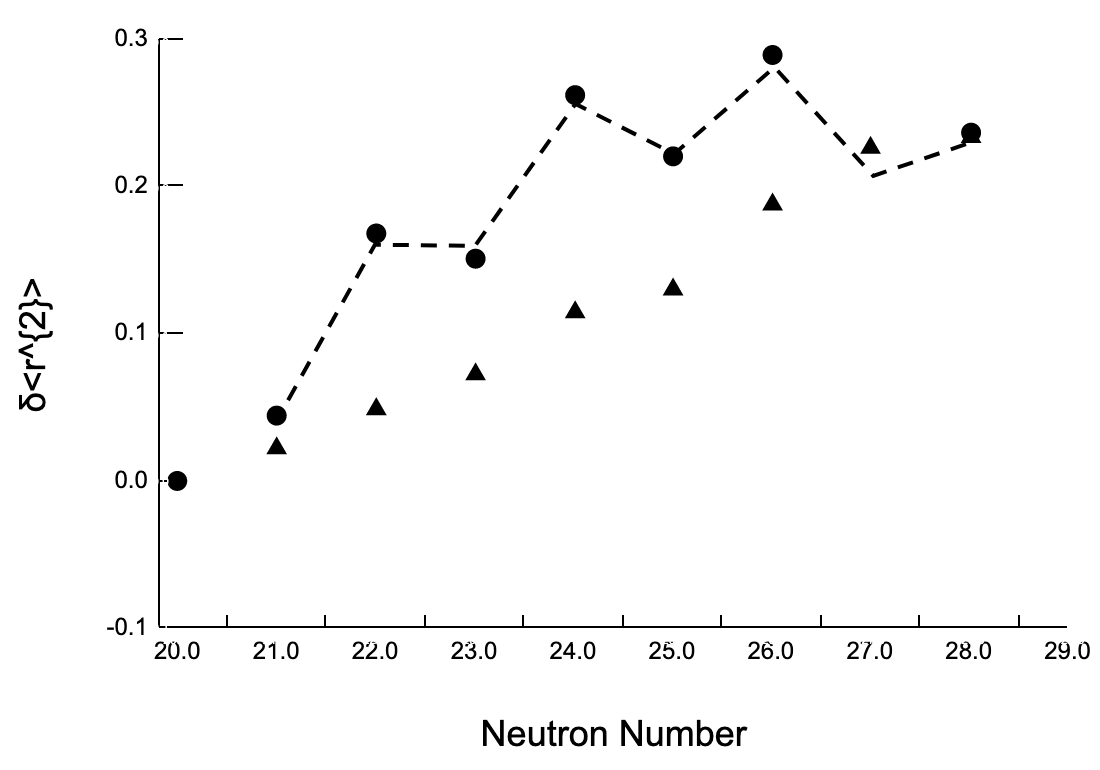}
        \caption{Isotope shifts in the odd $N$ Ar Isotopes. The filled circles correspond to the experiment of Blaum et al. \cite{2} and the dashed line the Zamick formula \cite{3}. The triangles correspond to spherical Hartee-Fock calculation \cite{5}\cite{6}}
        \label{fig:oAi}
    \end{figure}

    In Fig \ref{fig:oAi} we show measured values of isotope shifts in the Argon Isotopes by Blau et al \cite{2} (open circles). Also shown are spherical Hartree-Fock calculations in closed triangles, as well as a formula by Zamick \cite{3} and by Talmi \cite{5}\cite{6} which will soon be discussed.

    Note that the data shows a lot of even-odd staggering but the HF calculations do not. The Zamick-Talmi calculations yield excellent fits to the data and have the even-odd scattering features  well under control. In order to get the even-odd staggering Zamick \cite{3} introduced a 2 body effective radius operator in addition to the one body term.
    
    We simply make the assumption that the effective charge radius operator has a two-body part as well as one body part
        \begin{equation}
            \delta r_{\text{eff}}^{2} = \sum_{i}O(i) + \sum_{i<j}V(i,j)
        \end{equation}
     where the symbol $V$ for the two-body part has been written to suggest the similarity with the two-body potential, since both are scalars.
    
    The problem of evaluating this operator for n particles in the $j = f_{7/2}$ shell is exactly the same problem as calculating the binding energies of nuclei whose configuration consists of several nucleons in a single $j$ shell. This problem has been solved and used with great success by the ``Israeli group'' including de-Shalit, Racah, Talmi, Thieberger, and Unna \cite{4}. In analogy with their binding energy formula we get for the change in charge radius
        \begin{equation}
            \delta r^{2} (40 + n) = nC + \frac{n(n-1)}{2} \alpha + \left[ \frac{n}{2} \right]\beta,
        \end{equation}
    where
        \begin{equation}
            \left[ \frac{n}{2} \right] = 
            \begin{dcases}
                \frac{n}{2} & \quad \text{for even $n$}\\
                \frac{n-1}{2} & \quad \text{for odd $n$}.
            \end{dcases}
        \end{equation}
    The parameter $C$ comes from the one-body part and is equal to $\delta r^{2}(41)$, the difference in charge radius of $^{41}$Ca and $^{40}$Ca. The quantities $\alpha$ and $\beta$ come from the two-body part
        \begin{equation}
        \begin{split}
            \alpha &= -\frac{2(j+1)\bar{E}_{2} - E_{0}}{2j + 1},\\
            \beta &= \frac{2(j+1)(\bar{E}_{2} - E_{0})}{2j + 1},
        \end{split}
        \end{equation}
    where
        \begin{equation}
        \begin{split}
            E_{0} &= \langle j^{2}J = 0 | V | j^{2}J = 0 \rangle\\
            \bar{E}_{2} &= \frac{\sum_{J\neq 0}(2J+1) \langle j^{2}J | V | j^{2}J\rangle}{\sum_{J\neq 0}(2J+1)}.
        \end{split}
        \end{equation}

\section{Redmond Modifications and Counting Pairs}

    Akito Arima and Yu-Min Zhao have written many papers concerning relations of states in the single $j$ shell. One example is ``Number of States for Nucleons in a single $j$ shell'' \cite{7}. I was involved in this kind of business as well and was delighted by the generous references to my works by Arima and Zhao.
    
    Let me give one example ``New Relations for coefficients of fractional parentage'', where we simplify a recursion relation due to Redmond \cite{8}.
    
    \begin{figure}[H]
        \centering
        \captionsetup{width=0.98\linewidth}
        \caption{We quote from the paper \cite{8}:}
        \includegraphics[width=0.95\textwidth]{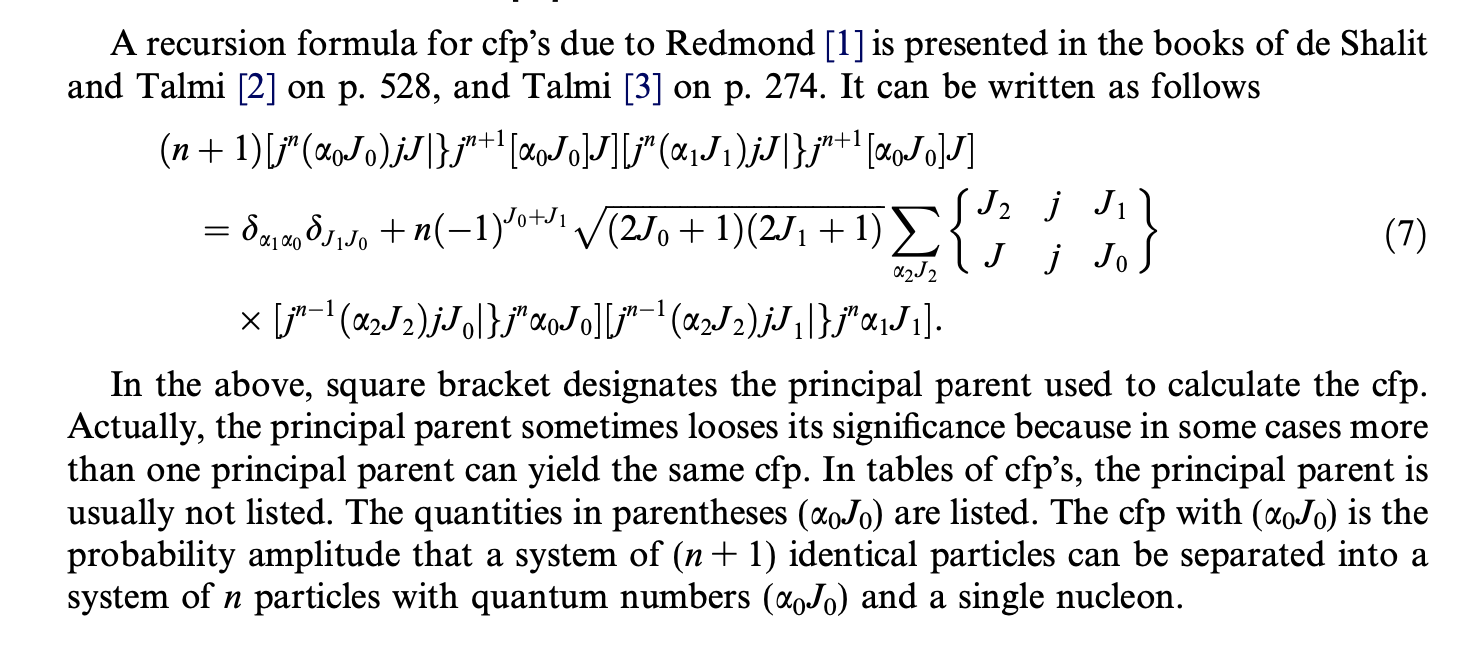}
        \label{fig:a1}
    \end{figure}
    
    \begin{figure}[H]
        \centering
        \includegraphics[width=0.95\textwidth]{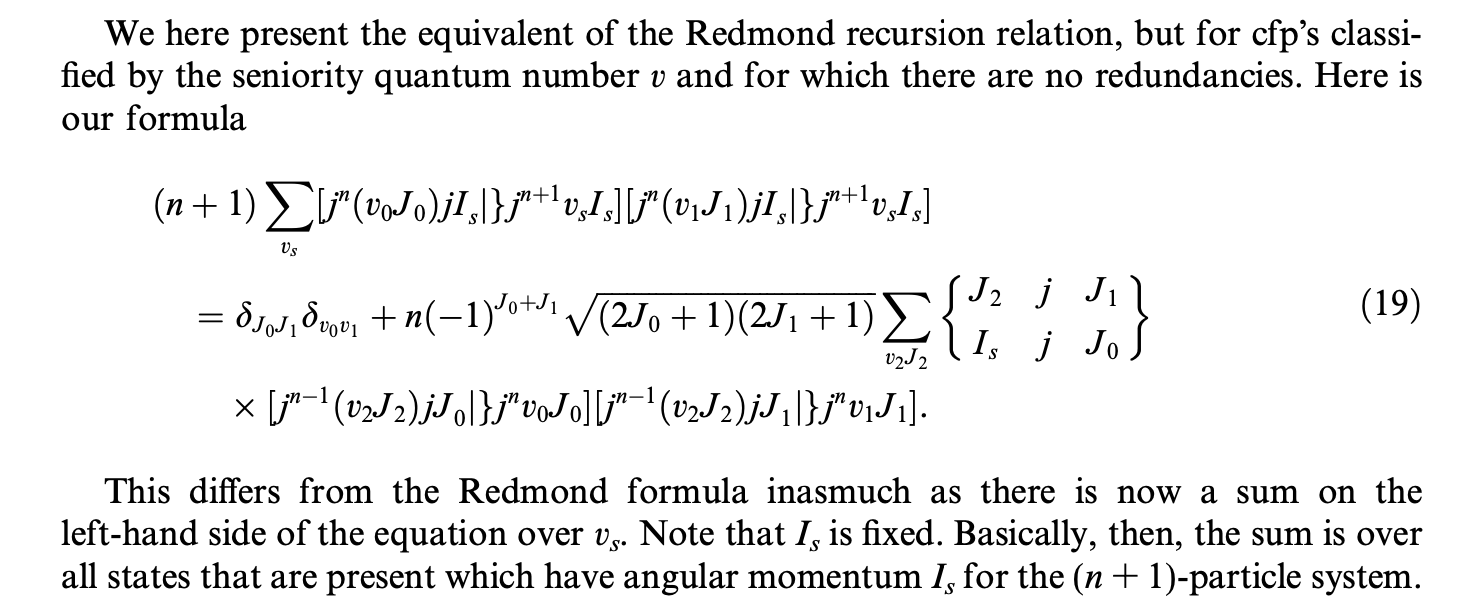}
        \label{fig:a2}
    \end{figure}

\section{Closing Remarks}

    Already in 1967, when Akito Arima was at Rutgers he was a big name on the world scene.Indeed when Gerry Brown went from Princeton to Stony Brook he took Arima with him as well as Tom Kuo. But as they say, the best was yet to come.I am sure the summary of all his accomplishments will appear somewhere in this compendium so I won't mention them. Well maybe a couple - pseudo spin and the interacting boson model. And in service-president of the University of Tokyo and head of Ricken. Rather I would like to dwell on the fun time it was to be a nuclear physicist in New Jersey around 1967. Princeton and Rutgers had a joint seminar called Nuclear News run by Rubby Sherr. Amongst the faculty, post-docs , senior grad students and visitors at that time were the following:
    \begin{itemize}
        \item Princeton: Gerry Brown, Tom Kuo, Tony Green, Chun WA Wong, George Bertsch, Felix Wong, Alex Lande Yitzhak Sharon, Julian Noble, Harry Mavromatis;
        \item Rutgers: Joe Ginocchio, Aldo Covello, Giovanni Sartoris, George Ripka and oh yes me.
    \end{itemize}
    And for icing on the cake. Shiro Yoshida, Akito Arima and Koichi Yazaki. Arima's discussions were appreciated not only by the theorists buy also the experimentalists - Noemie Koller, George Temmer, Rubby Sherr and others. Rutgers had a major program of measuring magnetic moments of excited states. Those were great times and Akito was a major contributor to the fun we all had.

\end{document}